\documentclass[%
 reprint,
 amsmath,amssymb,
 aps,
]{revtex4-2}
\usepackage{comment}
\usepackage{multirow}
\usepackage{hyperref}
\usepackage{nccmath}
\usepackage{graphicx}% Include figure files
\usepackage{dcolumn}% Align table columns on decimal point
\usepackage{bm}% bold math
\usepackage{color}

\usepackage{aas_macros}
\usepackage[T1]{fontenc}
\newcommand{\md}{\mathrm{d}}
\newcommand{\bb}[1]{\mathbf{#1}}

\begin{document}

\rightline{\scriptsize RBI-ThPhys-2024-10}

\preprint{APS/123-QED}

\title{COBRA: \\
Optimal Factorization of Cosmological Observables}% Force line breaks with \\

\author{Thomas Bakx}
 \email{t.j.m.bakx@uu.nl}
\affiliation{%
Institute for Theoretical Physics, \\
Utrecht University, \\
Princetonplein 5, 3584 CC, Utrecht,\\
The Netherlands.
}%
\author{Nora Elisa Chisari}
\affiliation{%
Institute for Theoretical Physics, \\
Utrecht University, \\
Princetonplein 5, 3584 CC, Utrecht,\\
The Netherlands.
}%
\author{Zvonimir Vlah}
\affiliation{Division of Theoretical Physics, Ruđer Bo\v{s}kovi\'c Institute, 10000 Zagreb, Croatia,}%
\affiliation{Kavli Institute for Cosmology, University of Cambridge, Cambridge CB3 0HA, UK}%
\affiliation{Department of Applied Mathematics and Theoretical Physics, University of Cambridge, Cambridge CB3 0WA, UK.}
\date{\today}% It is always \today, today,

\begin{abstract}
We introduce \texttt{COBRA} (Cosmology with Optimally factorized Bases for Rapid Approximation), a novel framework for rapid computation of large-scale structure observables. \texttt{COBRA} separates scale dependence from cosmological parameters in the linear matter power spectrum while also minimising the number of necessary basis terms $N_b$, thus enabling direct and efficient computation of derived and nonlinear observables. Moreover, the dependence on cosmological parameters is efficiently approximated using radial basis function interpolation. We apply our framework to decompose the linear matter power spectrum in the standard $\Lambda$CDM scenario, as well as by adding curvature, dynamical dark energy and massive neutrinos, covering all redshifts relevant for Stage IV surveys. With only a dozen basis terms $N_b$, \texttt{COBRA} reproduces exact Boltzmann solver calculations to $\sim 0.1\%$ precision, which improves further to $\sim 0.02\%$ in the pure $\Lambda$CDM scenario. Using our decomposition, we recast the one-loop redshift space galaxy power spectrum in a separable minimal-basis form, enabling $\sim 4000$ model evaluations per second at $\sim 0.02\%$ precision on a single thread. This constitutes a considerable improvement over previously existing methods (e.g., FFTLog) opening a new window for efficient computations of higher loop and higher order correlators involving multiple powers of the linear matter power spectra. The resulting factorisation can also be utilised in clustering, weak lensing and CMB analyses. Our implementation is publicly available at \url{https://github.com/ThomasBakx/cobra}. 
\end{abstract}

\maketitle

%\tableofcontents

\section{Introduction}
Large-scale structure (LSS) surveys mapping out the three-dimensional distribution of galaxies across billions of years of cosmic history will show us a unique imprint of the laws that govern our Universe. The Stage IV era of precision cosmology aims to probe the nature of dark matter and dark energy, the geometry of the Universe and the shape of its initial conditions of the first instance after the Big Bang \cite{desiv2,euclidv2,lsst}. As such, any tension with the baseline $\Lambda$CDM model (e.g., \cite{desibaov2}) could guide us to a deeper understanding of the answers to these fundamental questions. Correspondingly, the accuracy with which the distribution of galaxies and dark matter will be charted must be matched by higher accuracy of the corresponding theoretical model computation. 

Perturbation theory (PT) approaches to LSS \cite{bernardeaurev,schmidtrev,carrasco_fluid,baumann_eft} are a first-principle way of modelling the evolution of biased tracers of the dark matter density field such as galaxies. Correlators of biased tracers receive loop corrections that are expressable as integrals over the linear power spectrum $P_L(k)$. However, the dependence of even $P_L(k)$ on cosmological parameters is not analytically tractable, nor is the scale dependence for a given cosmology. Thus, direct implementation of these predictions (using Boltzmann solvers) is slow in any Bayesian approach, where likelihoods need to be sampled millions of times. This issue is exacerbated when considering higher-order corrections to summary statistics, which are integrals over $P_L(k)$. Conventional solutions fall into two classes. The first involves constructing an `analytical basis' of functions into which $P_L(k)$ is decomposed such that the resulting integrals can be evaluated exactly through tensor multiplications \cite{vlah_twoloop1,simonovic_fft,senatore_qcd,fft_pt}. The second approach is to emulate the resulting integrals as functions of scale and cosmology, via e.g. neural networks or other techniques \cite{bartlett,arico_emu,egge_emu,derose_emu,mancini_emu,cataneo_twoloop,chen_taylor,emu_trusov,ramirez_emu}. These approaches each have their drawbacks: \textit{first}, useful analytical bases are rare and typically do not approximate $P_L(k)$ well unless a large number of basis functions is used, which can lead to memory issues for higher-order statistics \cite{senatore_qcd,philcox_oneloopbisp}. Furthermore, the evaluation of the resulting tensors is still technically demanding and implementation is nontrivial, especially in redshift space \cite{damico_bisp}. Typically, techniques developed for a single observable at a specific perturbative order either lack efficient generalization to higher moments and higher perturbative orders or are rendered inapplicable altogether. In addition, such an approach still needs to be combined with a Boltzmann solver to compute $P_L(k)$ at a given cosmology. \textit{Second}, the emulation-based approach can require substantial computational resources and suffers from a lack of efficient generalization: every next quantity requires a sufficiently dense training set across all parameters.  \\

We pursue a different solution to this issue by finding an \textit{optimal factorization} of the scale dependence and cosmology dependence of the linear power spectrum. That is, we decompose it as
\begin{equation}\label{eq:decomp}
    P_L^\Theta(k) = \sum_{i=1}^{N_b} w_i(\Theta)v_i(k)
\end{equation}
where $\Theta$ indicates a set of cosmological parameters (including redshift). The $v_i(k)$ are fixed basis functions depending only on scale, which we call \textit{scale functions}. The \textit{weights} $w_i(\Theta)$ encode the cosmology dependence. Optimal factorization is achieved by choosing the {\it smallest} number of basis functions $N_b$ (see Section \ref{sec:method}). This decomposition allows for efficient calculation of higher-order statistics but does not rely on analytic methods for loop integrals nor a Boltzmann solver. We thus reap the benefits of both approaches while circumventing their shortcomings.

In Section \ref{sec:method}, we obtain such a decomposition and show that it facilitates computation of perturbative corrections. We then apply it to the $\Lambda$CDM $P_L(k)$ in Section \ref{sec:decomp}. As an illustrative example, in Section \ref{sec:rsd} we calculate the one-loop power spectrum of galaxies in redshift space rapidly and to high precision. We conclude in Section \ref{sec:disc}. Some technical aspects and extensions beyond $\Lambda$CDM are found in Appendices \ref{sec:rbf}, \ref{sec:ext} and \ref{sec:irres}. 

\section{Methodology}\label{sec:method}

Finding a set of scale functions that achieves a decomposition as in Eq.~\eqref{eq:decomp} amounts to finding a low-rank approximation of a set of \textit{template spectra} $P_{lm} = P_L^{\Theta_l}(k_m)$ evaluated at $N_t$ fixed cosmologies $\Theta_l$ and on a fixed set of wavenumbers $k_m$. This is achieved via a truncated singular value decomposition (SVD). Prior to performing the SVD, spectra are normalized by the mean of the templates $\bar{P}(k_m)$. %\footnote{In order for Eq. \eqref{eq:decomp} to remain satisfied, we can only apply \textit{linear and cosmology-independent} transformations to the templates - other commonly used transformations such as taking logarithms or dividing by a cosmology-dependent fitting formula would spoil the decomposition. We also attempted whitening the templates, i.e. subtracting the mean and dividing by the standard deviation, but this resulted in significantly degraded performance. }.
Writing $\hat{P}_{lm} = P_L^\Theta(k_m)/\bar{P}(k_m)$ and $\hat{v}_i(k_m) = v_i(k_m) / \bar{P}(k_m) $ we have
\begin{eqnarray}
    \hat{P} \approx \hat{U} \Sigma \hat{V}^T 
\end{eqnarray}
where $\hat{P}$ is $N_t \times N_k$, $\Sigma$ is diagonal and small ($N_b\times N_b$ where $N_b \ll N_k$), and $\hat{V}$ is $N_k \times N_b$ containing the principal components as its orthonormal column vectors, i.e. $\hat{V}_{mi} = \hat{v}_i(k_m)$. Lastly, $\hat{U}$ is $N_t \times N_b$ and contains the weights $w_i(\Theta_l)$ (for related work, see e.g. \cite{philcox_svd,pathak1,pathak2,arico_emu,derose_emu,mancini_emu}). 
%However, there are some important differences in our implementation.
%First, we do not opt for Latin Hypercube sampling across all parameters, but rather choose to densely sample the parameters that are expected to affect the shape of the power spectrum the most (see Section \ref{sec:decomp}). Second, the spectra used for the calculation of the weights $w_i(\Theta)$ are completely distinct from the templates used for computing the scale functions. 
Conducting the SVD is cheap \cite{halko} and can be done with many $(N_t > 10^7)$ templates, which need not be calculated exactly - they should only mimic the shape of $P_L(k)$ to ensure that Eq. \eqref{eq:decomp} is accurate.
The columns of $V$ span the \textit{optimal} $N_b$-dimensional approximation to the template set \cite{eytheorem}. The resulting scale functions are shown in Appendix \ref{sec:irres}, Figure \ref{fig:irres}.

Given scale functions, we compute weights via orthonormal projection:
\begin{eqnarray}\label{eq:proj}
    w_i(\Theta) = \sum_{m=1}^{N_k} \hat{v}_i(k_m)\hat{P}_L^\Theta(k_m).
\end{eqnarray}
We stress that evaluating the weights $w_i(\Theta)$ is a \textit{separate} problem, requiring either (i) exact calculation of $\hat{P}_L^\Theta(k_m)$ with e.g. \texttt{CAMB} and applying Eq. \eqref{eq:proj} or (ii) an indirect strategy using e.g. neural networks. We opt for a different indirect strategy based on \textit{radial basis functions} (RBFs) \cite{Buhmann_2003}, which we describe in Appendix \ref{sec:rbf}.

Armed with Eq. \eqref{eq:decomp} it becomes simple to compute next-to-leading order corrections to observables. For example, for the power spectrum (prior to IR-resummation) in redshift space at one-loop order (see e.g. \cite{senatore_bias,chen_velocileptors,senatore_redshift,perko_bias,angulo_bias}) one schematically has 
\begin{eqnarray}\label{eq:1loop1}
    P^\Theta_{\text{1-loop}}(k,\mu) &=& \text{const.}(k,\mu) + \mathcal{S}^{l}[P_L^\Theta](k,\mu)\nonumber \\ &+& \mathcal{S}^{q}[P^\Theta_L,P^\Theta_L](k,\mu)
\end{eqnarray}
where $\text{const.}(k,\mu)$ does not involve $P_L(k)$ while $\mathcal{S}^{l}$ and $\mathcal{S}^{q}$ are linear and quadratic operators that do not depend on cosmology. Here $l$ and $q$ superscripts refer to terms linear and quadratic in $P_L(k)$. Concretely, $\text{const.}(k,\mu)$ consists of stochastic terms while $\mathcal{S}^{l}$ involves the linear theory part and counterterms $\propto k^2 P_L(k)\mu^{2n}$, and finally, $\mathcal{S}^{q}$ consists of $(22)$ and $(13)$-type contributions to the loops. Plugging in Eq. \eqref{eq:decomp} yields
\begin{eqnarray}\label{eq:1loop2}
    P^\Theta_{\text{1-loop}}(k,\mu) &=& \text{const.}(k,\mu) + \mathcal{S}_i^{l}(k,\mu)w_i(\Theta)\nonumber \\ &+& \mathcal{S}_{ij}^{q}(k,\mu)w_i(\Theta)w_j(\Theta)
\end{eqnarray}
where $\mathcal{S}_i^{l} = \mathcal{S}^{l}[v_i]$ and $\mathcal{S}_{ij}^{q} = \mathcal{S}^{q}[v_i,v_j]$. This reduces calculating $P^\Theta_{\text{1-loop}}$ to multiplications of precomputed matrices whose entries are integrals of scale functions against PT kernels. Similar arguments apply to other N-point functions and higher PT orders \cite{simonovic_fft,senatore_qcd}. We can also extend this to the redshift-space galaxy power spectrum, including infrared (IR) resummation (see Appendix \ref{sec:irres}).

\section{Linear Power Spectrum}\label{sec:decomp}

We decompose of $P_L(k)$ in four scenarios, varying the cosmological parameter space ($\Lambda$CDM or \textit{generalized}) and ranges of parameters (\textit{default} or \textit{extended}). We choose the range $8 \times 10^{-4} h^*/\text{Mpc} < k < 4h^*/\text{Mpc}$. %The generalized cosmologies include massive neutrinos, curvature and dynamical dark energy. By extending the parameter ranges, we are able to gauge the dependence of the number of scale functions $N_b$ needed for a given level of precision on the range of cosmological parameters. The RBF interpolation is always done on the unit cube $[0,1]^D$, meaning that all quoted parameter ranges are linearly mapped onto this interval. Since our focus is on computing loop corrections to observables,  for $\Lambda$CDM and $10^{-3} h^*/\text{Mpc} < k < 1.5h^*/\text{Mpc}$ for the generalized cosmologies, but this is not a fundamental restriction \footnote{The reason for the distinct limits is the IR resummation prescription needed in the $\Lambda$CDM case for the one-loop power spectrum in redshift space. This requires smoothing the linear power spectrum, and we would like to maintain agreement in the IR-resummed power spectrum for $10^{-3} h^*/\text{Mpc} < k < 1.5h^*/\text{Mpc}$}.
We use $h^*=0.7$ and compute $P_L(k)$ with \texttt{CAMB} \cite{2011ascl.soft02026L,camb2} (v1.5.2). We show results for $\Lambda$CDM in the main text and defer generalized cosmologies including curvature, dynamical dark energy and neutrinos to Appendix \ref{sec:ext}. 

\begin{figure}[t]
\includegraphics[width=0.48\textwidth]{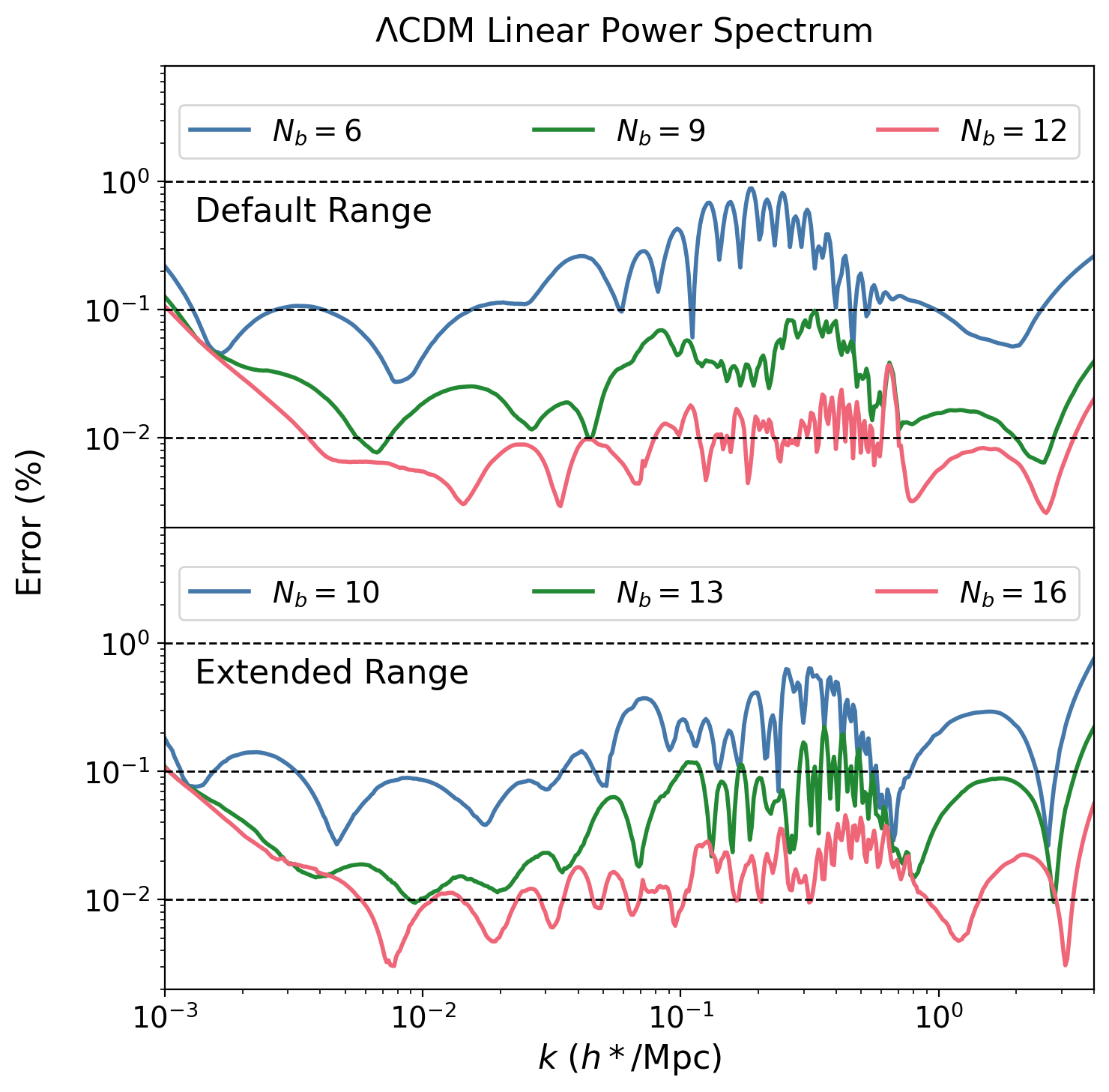}
\caption{\label{fig:plcdm} The $99.7$th percentile errors on the $\Lambda$CDM $P_L(k)$ for several choices of $N_b$, both for default (upper panel) and extended (lower panel) ranges. Dashed lines indicate $0.01\% \,, 0.1\%$ and $1\%$ errors, respectively.}
\end{figure}

In $\Lambda$CDM, we consider $\{\Theta\} = \{\omega_b, \omega_c, n_s, A_s,h,z\}$.
The shape of $P_L(k)$ does not depend on the evolution parameters $\Theta_e=\{A_s,h,z\}$ when the shape parameters $\Theta_s=\{\omega_c,\omega_b,n_s\}$ are held fixed. Thus, for the SVD we only vary $\omega_c,\omega_b$ and $n_s$ \cite{egge_emu,sanchez_evol1,sanchez_evol2}. The ranges of all parameters and choices for the SVD are indicated in Table \ref{tab:tab1} \footnote{We do not extend the range for $\omega_b$ appreciably since in the context of spectroscopic clustering one typically employs a BBN prior \cite{desibaov2}.}. We thus compute $P_L(k)$ at fixed evolution parameters $\Theta_e^*$ as
\begin{eqnarray}\label{eq:fixevo}
    P_L^{\Theta_s,\Theta_e^*}(k) = \sum_{i=1}^{N_b} w_i(\Theta_s)v_i(k)
\end{eqnarray}
and for arbitrary evolution parameters as
\begin{eqnarray}\label{eq:evol}
    P_L^{\Theta_s,\Theta_e}(k) = \frac{A_s}{A_s^*}\frac{D_+^2(\omega_m,h,z)}{D_+^2(\omega_m,h^*,z^*)}P_L^{\Theta_s,\Theta_e^*}(k) 
\end{eqnarray}
with $\omega_m = \omega_c+\omega_b$. The ratio of growth factors $D_+$ in Eq. \eqref{eq:evol} is also approximated using RBFs. We found it beneficial to first divide by the exact expression for a Universe with $\Omega_m + \Omega_\Lambda=1$.  \cite{vlah_twoloop2} %\footnote{There is a small difference between the exact growth factor for the $\Lambda$CDM cosmologies we consider here and the analytical expression for a Universe with $\Omega_m+\Omega_\Lambda=1$ due to nonzero radiation energy density - this is necessarily associated with a nonzero cosmic microwave background (CMB) temperature. For practical purposes, this difference is negligible, but we include it here for completeness.}. 
For the RBF approximations we use $N_n = 400$ Halton nodes \cite{owen2017randomized}. We use $5\,000$ cosmologies to test the precision of the predictions.
\begin{table}[t]
    \centering
    \begin{tabular}{|c|c|c|c|c|}
    \hline
         & \multicolumn{2}{c|}{Default} &
        \multicolumn{2}{c|}{Extended} \\
        \hline
        $\Theta $& Range & Grid size & Range & Grid size \\
        \hline\hline
        $\omega_c$ & [0.095,0.145] & 27 & [0.08,0.175] & 40 \\
        $\omega_b$ & [0.0202,0.0238] & 18 & [0.020,0.025] & 20\\
        $n_s$ & [0.91,1.01] & 12 & [0.8,1.2] & 20 \\
        $10^9 A_s$ & - & $10^9 A_s^* = 2$ & - &$10^9 A_s^* = 2$ \\
        $h$ & [0.55,0.8] & $h^* = 0.7$ & [0.5,0.9] & $h^* = 0.7$\\
        $z$ & [0.1,3] & $z^* = 0$ & [0.1,3] & $z^* = 0$\\
        \hline
    \end{tabular}
    \caption{Ranges and (linearly spaced) template grids for $\Lambda$CDM parameters. If a parameter is held fixed, its fiducial value is indicated. %For example, for the default parameter range we used a grid of size $27 \times 18 \times 12 = 5\,832$ in $\omega_c,\omega_b$ and $n_s$, keeping $A_s,h$ and $z$ fixed.
    }
    \label{tab:tab1}
\end{table}
The result is shown in Figure \ref{fig:plcdm}. 
The default (extended) range requires $N_b=9\, (13)$ basis functions for $0.01\%$ precision for $99.7\%$ ($3\sigma$) of the test cosmologies. %The error in Figure \ref{fig:plcdm} is determined mostly by the number of basis functions used and the accuracy of the power spectrum factorization from Eq. \eqref{eq:evol}\footnote{This factorization is in fact not exact; there is a small residual scale dependence at large scales which explains the upturn in Figure \ref{fig:plcdm} for small $k$}.
Increasing $N_b$ to $12\,(16)$ decreases the $3\sigma$ error to $\sim 0.02\%$ \footnote{While the agreement between the different Boltzmann codes \texttt{CAMB} and \texttt{CLASS} \cite{class_approx} may not be at that level \cite{camb_class,arico_emu}, it is a testament to the precision of our method that such small errors can be achieved.}. One prediction for $P_L(k)$ takes $\sim 0.4$ ms, while vectorized evaluation yields $250$ spectra in $\sim 4$ ms, all on one thread \footnote{Tests are run on an Apple M1 Pro processor (16GB RAM).}. %The dependence of the timings on $N_b$ is very mild. 
\section{One-Loop Galaxy Power Spectrum}\label{sec:rsd}
Using the decomposition of $P_L(k)$ from Section \ref{sec:decomp}, the calculation of higher-order corrections to N-point functions is straightforward. We illustrate this using the one-loop power spectrum of galaxies in redshift space for a $\Lambda$CDM cosmology, but emphasize that this choice is irrelevant - the calculation of higher N-point functions requires only a one-time computation of a limited number of integrals which can be done using \textit{any method}, regardless of whether analytical techniques are available. 
\begin{figure}[t]
\includegraphics[width=0.48\textwidth]
{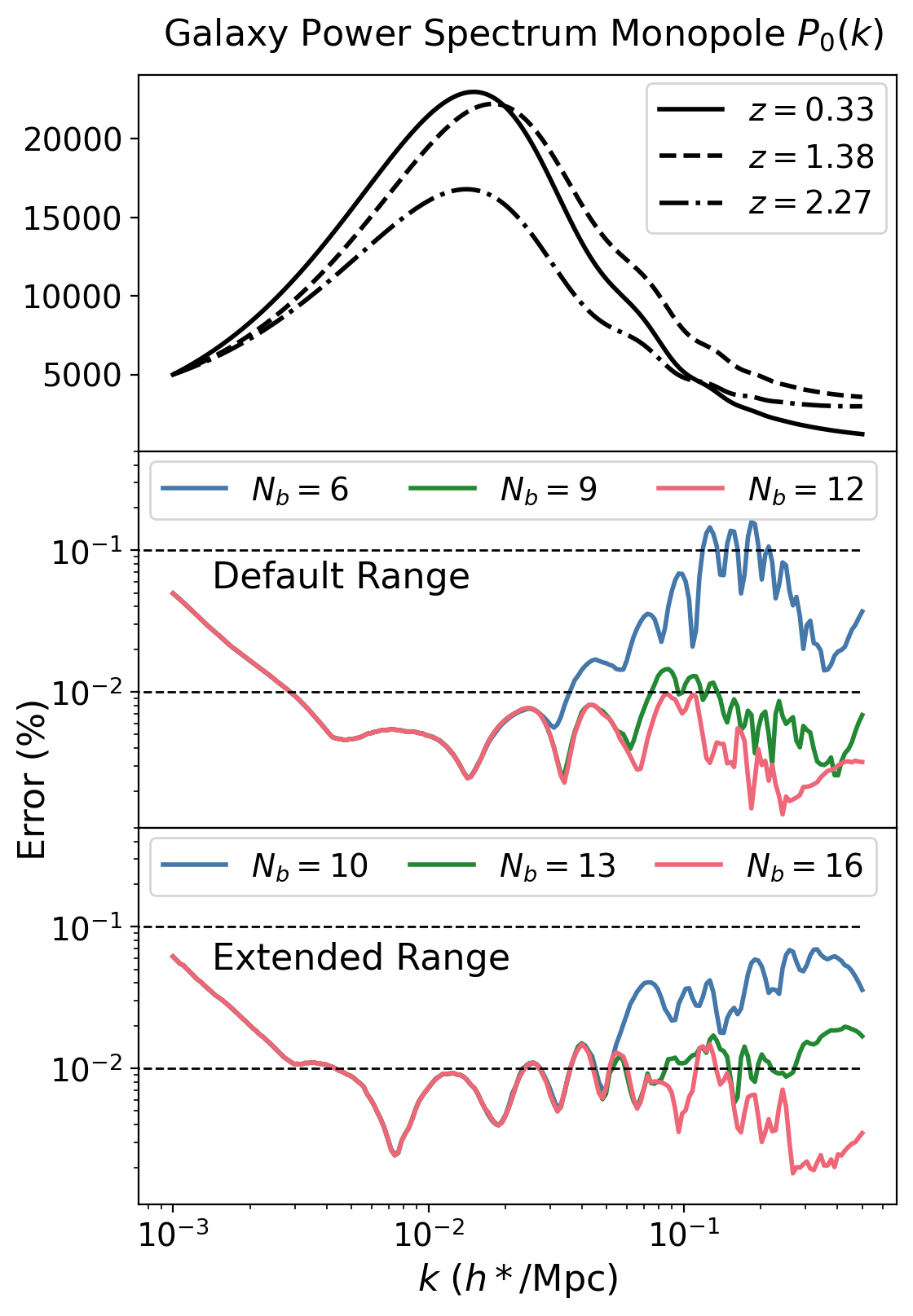}
\caption{\label{fig:p1loop} Performance of \texttt{COBRA} for the monopole of the IR-resummed one-loop power spectrum of galaxies in redshift space. The top panel shows three random cosmologies (solid, dashed and dot-dashed lines) and their redshifts. The second and third panels show the $99.7$th percentile of the errors on the test set for the default and extended ranges. For visualization, spectra are normalized to the same overall amplitude.}
\end{figure}

To test \texttt{COBRA}, we compare against the one-loop model implemented in \texttt{velocileptors} \cite{chen_velocileptors}. All terms in this model are either constant, linear or quadratic functions of $P_L^\Theta$ (cf. Eq. \ref{eq:1loop1}), so that by Eq. \eqref{eq:1loop2} they reduce to matrix multiplications of size $N_b$. %when Eq. \eqref{eq:decomp} is substituted for $P_L(k)$. 
%For $\mathcal{S}_i^l$ we use $N_b=12\, (16)$ for the linear part in the default (extended) range and assess various choices for $N_b$ when calculating $S_{ij}^q$. 
We employ the same parameter setup as in Section \ref{sec:decomp}. %We apply a Gaussian damping factor to the scale functions $v_i(k)$ to extrapolate beyond $k=4h^*$/Mpc, and a power law at low $k$. 
%Since we compute loop integrals only once, we choose a large number of frequencies $N_f = 10\,000$. 
The IR resummation prescription is detailed in Appendix \ref{sec:irres}. We keep operator biases fixed to the values listed in \footnote{\url{https://github.com/sfschen/velocileptors/blob/master/notebooks/EPT\%20Examples.ipynb}}. We fix $A_s=2 \times 10^{-9}$ and put all coefficients $\propto k^2\mu^{2n}P_{L,\text{IR}}(k)$ equal to $40 \,(50)$ in the default (extended) case and counterterms $\propto k^2\mu^{2n}$ to $3\,000$ \footnote{We do this purely to avoid zero crossings in the monopole, which are unphysical but nevertheless could occur due to large loop contributions for some cosmologies.}. The $k \to 0$ contributions are subtracted to recover linear theory on large scales. We use RBF approximations for the $\Lambda$CDM growth rate and velocity dispersion (see Appendix \ref{sec:irres}); their impact on the error is small. We omit Alcock-Paczynski rescaling, but this can be included at no cost since \texttt{COBRA} computes the full anisotropic power. 

Figure \ref{fig:p1loop} displays the error for the monopole using \texttt{COBRA} versus using \texttt{velocileptors} for $ 10^{-3} h^*/\text{Mpc} < k < 0.5h^*/\text{Mpc}$, using the same test set as in Section \ref{sec:decomp}. We obtain similar results for the quadrupole and hexadecapole. For the default (extended) range, we use $N_b = 12\, (16)$ for the linear part $\mathcal{S}_i^l$ from Eq. \eqref{eq:1loop2}. %hence the overlapping error bands on linear scales. 
We explore different choices of $N_b$ for $\mathcal{S}^q_{ij}$, which dominates computation time. For the default (extended) range, using $N_b = 9 \, (13)$ scale functions for $\mathcal{S}^q_{ij}$ we reach $\sim 0.01\%$ precision for $99.7\%$ of all test cosmologies. With $N_b = 12 \, (16)$ for $\mathcal{S}^q_{ij}$ and $200$ $k$-bins, the matrices needed for all bias terms require around $25\,(40)$ MB memory. One prediction of three multipoles takes $\sim 2.5$ ms, while $250$ predictions take $\sim 65$ ms. Thus, \texttt{COBRA} executes $\sim 4\,000$ predictions per second. This speed is unaltered when varying bias parameters. 

\section{Discussion}\label{sec:disc}
We introduced \texttt{COBRA}, a method for efficient computation of large-scale structure observables, and applied it to the linear power spectrum and the one-loop power spectrum of galaxies in redshift space. Generalising from Section \ref{sec:method} (and ignoring IR resummation for simplicity, see Appendix \ref{sec:irres}), all polyspectra $\mathcal{P}$ take the form \cite{simonovic_fft,senatore_qcd}
\begin{eqnarray}\label{eq:genexp}
    \mathcal{P}^\Theta_{\ell\text{-loop}} &=& \text{const.} + \mathcal{S}^{l}[P_L^\Theta] + \mathcal{S}^{q}[P^\Theta_L,P^\Theta_L] \\
    &+& \mathcal{S}^{c}[P^\Theta_L,P^\Theta_L,P^\Theta_L]
    + \dots \nonumber 
\end{eqnarray}
where $\mathcal{S}^l,\mathcal{S}^q$ and $\mathcal{S}^c$ are linear, quadratic and cubic operators \textit{et cetera}. %All operators are symmetric in their arguments, even in the case of cross-correlations of different tracers. % Indeed, the cross-correlation of different tracers only affects the functional correlator dependence via different combinations of bias coefficients. 
Using Eq. \eqref{eq:decomp}, 
\begin{eqnarray}\label{eq:tensordec}
    \mathcal{P}^\Theta_{\ell \text{-loop}} &=& \text{const.} + \mathcal{S}_i^{l}w_i(\Theta) + \mathcal{S}_{ij}^{q}w_i(\Theta)w_j(\Theta) \nonumber \\
    &+& \mathcal{S}_{ijk}^{c}w_i(\Theta)w_j(\Theta)w_k(\Theta) + \dots 
\end{eqnarray}
with $\mathcal{S}_i^l = \mathcal{S}^l[v_i],\mathcal{S}_{ij}^q = \mathcal{S}^q[v_i,v_j]$ and $\mathcal{S}_{ijk}^c = \mathcal{S}^c[v_i,v_j,v_k]$ symmetric tensors. %Thus, for the $d$-th order term in Eq. \eqref{eq:genexp} (with $d$ arguments of $P_L$), the total number of tensor elements that need to be computed in a per wavenumber bin and per bias term is $(N_b+d-1)!/(d!(N_b-1)!)$.
Most terms contain factors of $P_L(k)$ outside the loop integral. For the one-loop bispectrum, the most demanding contribution is the $B_{222}$ diagram which integrates over three $P_L(k)$. Na\"ive counting with $N_b = 6$ and $15$ bias terms yields $\sim1\,000$ integrals to be evaluated, which is much more efficient than traditional methods \cite{simonovic_fft,philcox_oneloopbisp}. Moreover, the decay of singular values in SVD employed in \texttt{COBRA} (see Appendix \ref{sec:irres}) implies that the number of basis functions necessary to reach a given precision can be further reduced in terms that involve products of $P_L(k)$. Integrated quantities (like mode coupling terms) might benefit from additional reduction in the number of terms without compromising precision \cite{senatore_qcd}. Accuracy requirements also loosen substantially when data covariance is taken into account \cite{emu_trusov,class_oneloop}. Given the small number of basis functions needed, there is ample opportunity to generalize our calculations to e.g. exact time dependence \cite{vlah_twoloop2,choustikov,donath_exact,garny}, scale-dependent growth \cite{levi} or higher-order N-point functions \cite{egge_oneloopbisp,egge_oneloopbisp2,damico_bisp,damico_bisp2,donath,osato_twoloop1,osato_twoloop2,baldauf_oneloopbisp,baldauf_twoloopbisp,baldauf_onelooptrisp,senatore_twoloop,bertolini_onelooptrisp}.

To summarize, \texttt{COBRA} is: 
\paragraph{fast:} The calculation for $P_L(k)$ takes $\sim 1$ ms on one thread while the loop calculation takes $\sim 2$ ms. Implementations are vectorized, which implies further speed-ups when computing $\mathcal{O}(100)$ predictions simultaneously. This can be exploited in conjunction with vectorized likelihood samplers \cite{ultranest}.
\paragraph{precise:} We reach $ \sim 0.1\%$ precision on all observables considered. Precision can be adjusted by tuning $N_b$, which is useful if further speed-up is desired and does not require recalculating the tensors in Eq. \eqref{eq:tensordec}. 
\paragraph{general:} Analytical tools for calculating loop corrections are unnecessary and our method applies to any N-point function at any loop order.  
\paragraph{lightweight:} The tabulated scale functions and tensors for the one-loop integrals require negligible memory. Our is publicly available at \url{https://github.com/ThomasBakx/cobra} and requires only \texttt{numpy} and \texttt{scipy}. 

No other method available in the literature can produce one-loop predictions with such efficiency and precision. {\tt COBRA} is applicable to all orders and higher-loop computations, making it a more versatile and powerful tool as well. In particular, it would be interesting to explicitly consider the calculation of higher N-point functions such as the one-loop bispectrum and two-loop power spectrum - improvements over existing methods will be even more pronounced here. The small number of coefficients used to approximate the $P_L(k)$ facilitates direct reconstruction \cite{amendola1,amendola2,freepower}. Other cosmology-independent and linear operations on observables, such as window convolution \cite{pardede_window} or compression schemes \cite{philcox_svd} could be incorporated in \texttt{COBRA} as a one-time calculation at the level of scale functions. One could explore taking analytic derivatives of RBF approximations in the context of gradient-based sampling methods \cite{jaxcosmo,bonici}. Finally, while our focus in terms of scales and parameter ranges has been on 3D galaxy clustering on large scales, it would be valuable to extend this to sky-projected \cite{Gao:2023, Raccanelli:2023} or nonlinear quantities. These are more relevant for weak lensing and CMB probes.
\begin{acknowledgments}
We thank David Alonso, Marco Bonici, Giovanni Cabass, Alexander Eggemeier and Pedro Ferreira for useful comments. We acknowledge extensive use of the open-source Python libraries \texttt{numpy}, \texttt{scipy} and \texttt{scikit-learn}.

This publication is part of the project ``A rising tide: Galaxy intrinsic alignments as a new probe of cosmology and galaxy evolution'' (with project number VI.Vidi.203.011) of the Talent programme Vidi which is (partly) financed by the Dutch Research Council (NWO). For the purpose of open access, a CC BY public copyright license is applied to any Author Accepted Manuscript version arising from this submission.  Z.V. acknowledges the support of the Kavli Foundation.
\end{acknowledgments}

%\nocite{*}
\bibliography{apssamp}% Produces the bibliography via BibTeX.

\newpage
\appendix

\setcounter{page}{1}

\section{Radial Basis Functions}\label{sec:rbf}
Our strategy for calculating $w_i(\Theta)$ is based on the theory of \textit{radial basis functions} (RBFs) \cite{Buhmann_2003}. RBFs approximate a multivariate function $g(\Theta)$ in terms of radial one-variable functions $\phi(\theta)$ where $\theta = |\Theta|$ is the length of the $D$-dimensional vector $\Theta$. We will assume without loss of generality that $\Theta \in [0,1]^D$ - we can always achieve this by applying a linear transformation to the parameter ranges. The idea is simple: we center the radially symmetric functions at a predetermined set of $N_n$ nodes and demand that the interpolant matches the function at the nodes. 

More formally, once the nodes $\{\Theta_a\}_{1\leq a \leq N_n}$ are chosen we define the $N_n \times N_n$ interpolation matrix $\Phi$ with entries 
\begin{equation}\label{eq:intmatrix}
    \Phi_{ab} = \phi(|\Theta_a - \Theta_b|) = \phi_b(\Theta_a),
\end{equation}
i.e. $\phi_b(\Theta) = \phi(|\Theta-\Theta_b|)$. We choose the nodes to be a scrambled Halton set of suitable size \cite{owen2017randomized}, depending on the dimensionality of the parameter space (but more common choices such as a Latin hypercube would likely also suffice). Then, an approximation to $g$ is constructed by demanding that 
\begin{equation}\label{eq:interp}
    g(\Theta_a) = \sum_{b=1}^{N_n} c_b \Phi_{ab}
\end{equation}
and solving the linear system for the coefficients $c_b$. The interpolant is thus given by 
\begin{equation}\label{eq:interpolant}
    g(\Theta) = \sum_{b=1}^{N_n} c_b \phi_b(\Theta).
\end{equation}
We will find that at most $N_n \sim 10^4$ is sufficient (and these function values of course need to be calculated exactly), which is indeed much less than the number of templates used for the SVD. 

\begin{figure}[t]
\includegraphics[width=0.48\textwidth]
{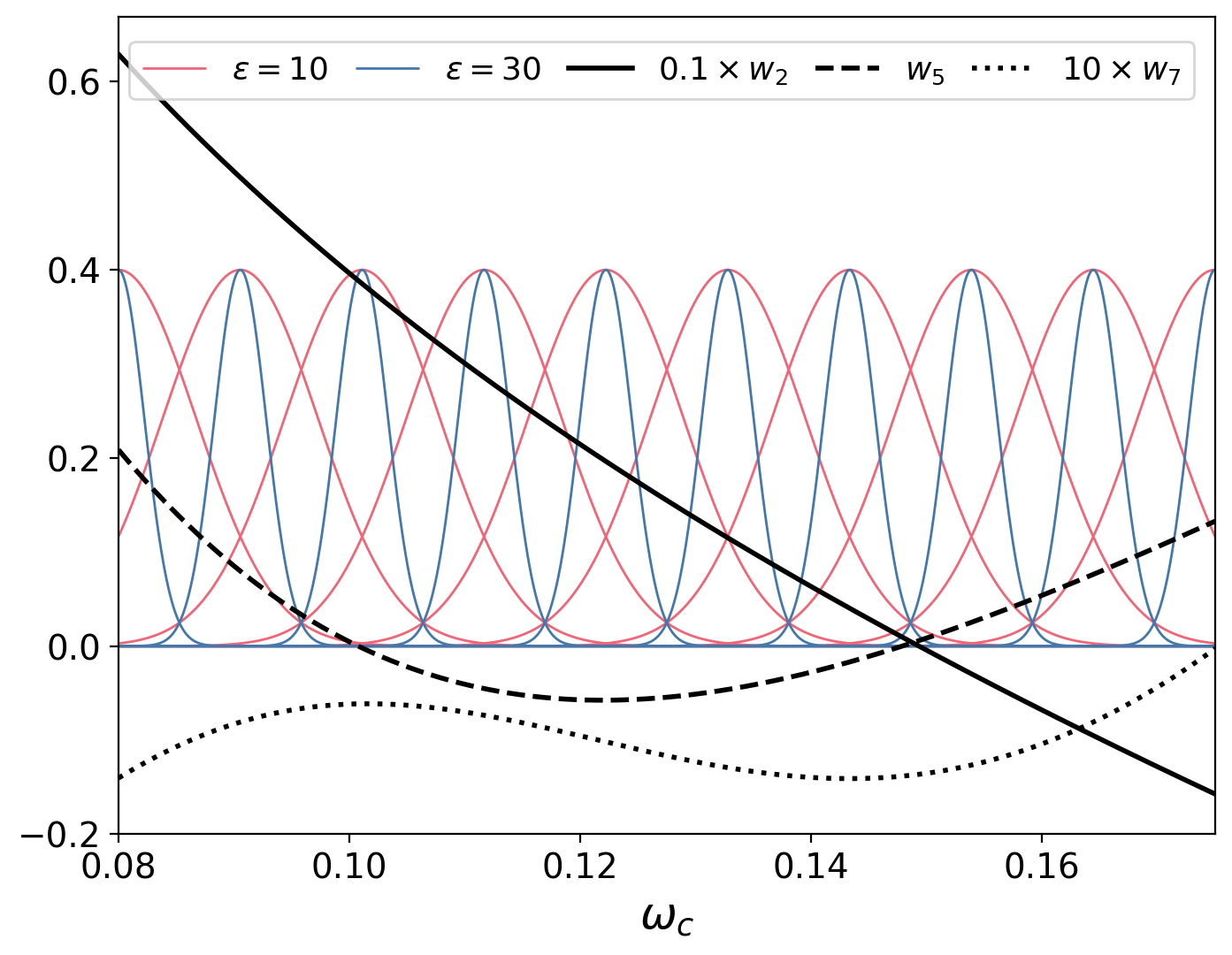}
\caption{\label{fig:rbffig}  An illustration of the RBF method for a one-dimensional slice of the 3D weight functions $w_i(\Theta)$ for the extended $\Lambda$CDM space, where $\omega_c$ is varied while $\omega_b$ and $n_s$ are held fixed. In pink and blue, we show some examples of RBF basis functions centred at 10 equally spaced values between $\omega_c = 0.08$ and $\omega_c = 0.175$ (but note that these do not match the Halton nodes we choose for the actual interpolation). Decreasing $\epsilon$ results in flatter Gaussians. The black lines indicate some of the weight functions, notably $w_2$, $w_5$ and $w_7$.
Observe that all curves except for $w_5$ have been rescaled in overall height for purely visual purposes.}
\end{figure}

A commonly used choice for $\phi$ is the \textit{Gaussian kernel}: 
\begin{equation}\label{eq:gauss}
        \phi^\epsilon(\theta) = \exp(-\epsilon^2 \theta^2).
\end{equation}
Here $\epsilon$ is an inverse length scale, or \textit{shape parameter}, which characterizes the width of the Gaussian and can be tuned for the application at hand \footnote{Here and throughout, we will use \textit{isotropic} Gaussians. It would certainly be interesting to consider generalizations with different (physically motivated) values of $\epsilon$ in different dimensions - this is mathematically straightforward \cite{fasshauer} and could lead to improved performance with fewer nodes.}. Figure \ref{fig:rbffig} illustrates this idea for a specific slice of the $\Lambda$CDM weight functions. 
The problem of finding an optimal shape parameter is highly relevant and optimal choices can lead to several orders of magnitude improvement in the error of the resulting interpolant. Specifically, if the shape parameter is too large, the basis functions do not have enough support away from the nodes and the quality of the approximation is poor. Conversely, a small choice of shape parameter (say, $\epsilon < 1$) can give extremely accurate results \cite{Fornberg2011StableCW,fasshauer,rbf_sphere,flatlimit_rbf}. The implementation of such an approach is however not trivial due to numerical instabilities that arise when attempting to compute $g(\Theta)$ via Eq. \eqref{eq:interpolant}. This is because the system in Eq. \eqref{eq:interp} is ill-conditioned for small $\epsilon$. The intuition is straightforward: when $\epsilon$ is small, the basis functions become increasingly flat and the columns of $\Phi_{ab}$ become degenerate. Additionally, large cancellations occur when computing the interpolant via Eq. \eqref{eq:interpolant} since the coefficients $c_b$ change sign often. However, \textit{the resulting interpolant depends smoothly on the parameter $\epsilon$} (see e.g. \cite{WRIGHT2017137, Fornberg2011StableCW,rbf_sphere,flatlimit_rbf} and references therein). In other words, it is possible to compute the interpolant in a numerically stable way \cite{fasshauer} (which we explain below), just not by na\"ively applying the two ill-conditioned steps of computing $c_b$ from Eq. \eqref{eq:interp} and the sum in Eq. \eqref{eq:interpolant} in succession. 

\begin{figure}[t]
\includegraphics[width=0.48\textwidth]
{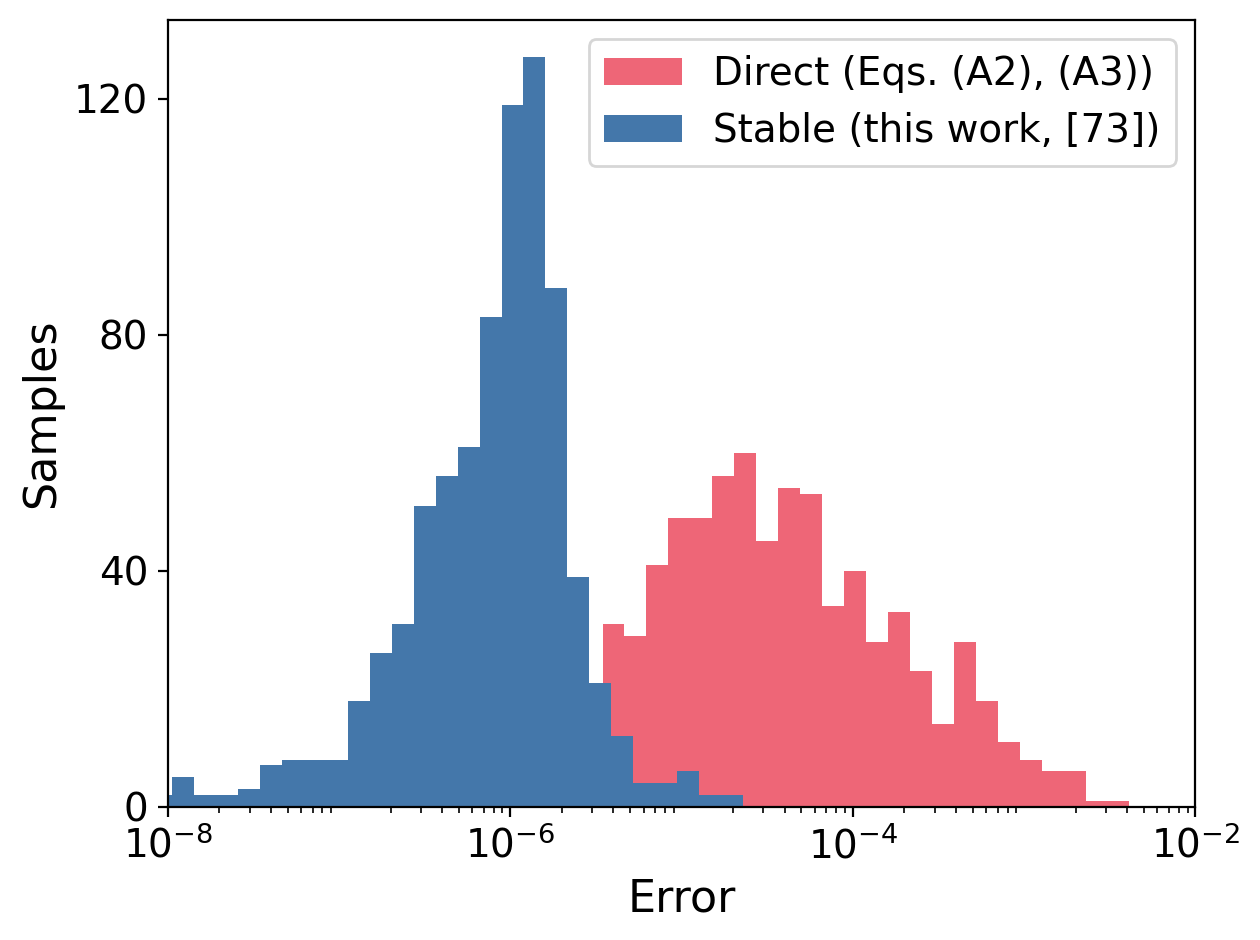}
\caption{\label{fig:rbfhist} Histogram of absolute relative errors made for $w_1(\Theta_s)$ from Eq. \eqref{eq:fixevo} in the extended $\Lambda$CDM parameter space. We used $N_t = 800$ test cosmologies and $N_n = 400$ Halton nodes. For the direct approach (pink), we used $\epsilon = 5$, in which case the condition number of the interpolation matrix $\Phi$ from Eq. \eqref{eq:intmatrix} already exceeds $10^{11}$. Employing the stable approach (blue) from \cite{fasshauer} with $\epsilon = 0.1$ and $\alpha = 1.8$ reduces the largest incurred error by more than two orders of magnitude.}
\end{figure}

This is illustrated in Figure \ref{fig:rbfhist} where we show the result of applying both the direct (i.e. with Eqs. \eqref{eq:interp}, \eqref{eq:interpolant})) and the stable RBF interpolation with $N_n = 400$ nodes for the first $\Lambda$CDM weight $w_1(\Theta_s) = w_1(\omega_c,\omega_b,n_s)$ (cf. Eq. \eqref{eq:fixevo}). For the direct approach we used $\epsilon = 5$, which results in a condition number for $\Phi_{ab}$ of roughly $3 \times 10^{11}$. When attempting to decrease $\epsilon$ further, the condition number of $\Phi$ quickly exceeds $10^{16}$ and thus the problem becomes unsolvable in standard 16-bit floating precision arithmetic. However, the stable implementation is able to cross this `barrier' and achieve much smaller errors. Notably, the maximum incurred error for the stable approach is roughly $2 \times 10^{-5}$, more than two orders of magnitude smaller than the largest error in the direct approach. We obtain similar results for higher weights $w_i(\Theta)$ with $i>1$, although errors are artificially larger due to zero crossings.

We now explain the details of the numerically stable procedure that realizes the calculation of $g(\Theta)$ for small $\epsilon$, following \cite{fasshauer} and focusing first on $D=1$ for simplicity. First, we consider the expansion 
\begin{equation}\label{eq:mercer}
    e^{-\epsilon^2(\theta - \theta')^2} \approx \sum_{k=1}^M \lambda_k^\epsilon \varphi_k^\epsilon(\theta) \varphi_k^\epsilon(\theta')
\end{equation}
(which is exact when $M\to \infty$) where the functions $\varphi_k^\epsilon(\theta)$  are given by 
\begin{equation}\label{eq:herm}
    \varphi_k^\epsilon(\theta) = \sqrt{\frac{\beta}{2^{k-1}(k-1)!}}e^{-\delta^2\theta^2}H_{k-1}(\alpha\beta\theta)
\end{equation}
with $\alpha$ a free hyperparameter whose value we choose below, $\beta^2 = \sqrt{1+(2\epsilon/\alpha)^2}$ and  $\delta^2=\alpha^2(\beta^2-1)/2$. Furthermore $H_{k-1}$ is a Hermite polynomial:
\begin{equation}
    H_k(x) = (-1)^k e^{x^2}\frac{\md^k}{\md x^k}e^{-x^2}.
\end{equation}
The sequence $\lambda_k^\epsilon$ is exponentially decaying and given by
\begin{equation}
    \lambda_k^\epsilon = \sqrt{\frac{\alpha^2}{\alpha^2+\epsilon^2+\delta^2}}\bigg( \frac{\epsilon^2}{\alpha^2+\epsilon^2+\delta^2}\bigg)^{k-1}.
\end{equation}
The $\varphi_k^\epsilon(\theta)$ are orthonormal on the real axis with respect to a Gaussian weight function: 
\begin{equation}
    \int_{-\infty}^\infty \md \theta \frac{\alpha}{\sqrt{\pi}}e^{-\alpha^2 \theta^2} \varphi_k^\epsilon(\theta)\varphi_{k'}^\epsilon(\theta) = \delta_{kk'};
\end{equation} 
by virtue of the standard orthogonality property of the Hermite polynomials \footnote{In fact, when $\epsilon \to 0$ the $\varphi_k^\epsilon(\theta)$ reduce to Hermite polynomials; polynomial regression techniques were also considered by \cite{knabe1,knabe2_v2} for the \texttt{EuclidEmulator}.}. This identity shows that $\alpha$ controls the scale over which the functions $\varphi_k^\epsilon$ are supported. Plugging Eqs. \eqref{eq:gauss}, \eqref{eq:mercer} into Eq. \eqref{eq:intmatrix} we get
\begin{equation}\label{eq:lowrank}
    \Phi = (\tilde{\Phi}^\epsilon)\Lambda^\epsilon(\tilde{\Phi}^{\epsilon})^T 
\end{equation}
where $\Lambda^\epsilon$ is an $M \times M$ diagonal matrix with $\lambda_k^\epsilon$ on the diagonal and $\tilde{\Phi}^\epsilon$ is $N_n \times M$ %i.e. `slim', since $M < N_n$ in practice, 
with entries $\tilde{\Phi}_{ak}^\epsilon = \varphi_k^\epsilon(\theta_a)$. If we choose $M<N_n$, Eq. \eqref{eq:lowrank} is effectively a low-rank approximation of $\Phi_{ab}$.  %Early truncation of the expansion from Eq. \eqref{eq:mercer} has the effect of making the calculation more stable. 
%If the value of $\alpha$ is chosen suitably, the matrix $\tilde{\Phi}^\epsilon$ is much better conditioned than $\Phi^\epsilon$. It admits a \textit{pseudo-inverse} $(\tilde{\Phi}^\epsilon)^\dagger$ which can be computed via QR-decomposition without catastrophic loss of precision. 
The fact that $\Phi$ is nearly singular is reflected in the fact that the entries of $\Lambda^\epsilon$ are small when $\epsilon$ is small. By contrast, the matrix $\tilde{\Phi}^\epsilon$ is much better conditioned. It admits a \textit{pseudo-inverse} $(\tilde{\Phi}^\epsilon)^\dagger$, which satisfies 
\begin{eqnarray}
    (\tilde{\Phi}^\epsilon)^\dagger \tilde{\Phi}^\epsilon = \mathbb{I}_M
\end{eqnarray}
where $\mathbb{I}_M$ is the $M\times M$ identity matrix. 

We now show how to compute the interpolant $g(\theta)$ at an arbitrary value of $\theta$ using Eq. \eqref{eq:mercer} and Eq. \eqref{eq:lowrank}. Using $\tilde{\Phi}^\epsilon(\theta)$ as a shorthand for the $M$-dimensional vector $\varphi_k^\epsilon(\theta)$ and plugging Eq. \eqref{eq:mercer} into Eq. \eqref{eq:interpolant} we get the following expression for the interpolant:
\begin{eqnarray}\label{eq:gtheta}
    g(\theta) = \tilde{\Phi}^\epsilon(\theta) \Lambda^\epsilon(\tilde{\Phi}^{\epsilon})^T \hat{c}
\end{eqnarray}
where $\hat{c}=c_b$ is the vector of coefficients. We can also rewrite Eq. \eqref{eq:interp} by using Eq. \eqref{eq:lowrank}. Introducing the vector $\hat{g} = g(\Theta_a)$, we obtain 
\begin{eqnarray}\label{eq:ghat}
    &\hat{g} = \Phi\, \hat{c} = (\tilde{\Phi}^\epsilon)\Lambda^\epsilon(\tilde{\Phi}^{\epsilon})^T \hat{c} \nonumber \\
    \implies & (\tilde{\Phi}^\epsilon)^\dagger \hat{g} = \Lambda^\epsilon(\tilde{\Phi}^{\epsilon})^T \hat{c}.
\end{eqnarray} 
Finally, after plugging Eq. \eqref{eq:ghat} into Eq. \eqref{eq:gtheta}, we obtain an expression for $g(\theta)$ which \textit{does not involve $\Lambda^\epsilon$}. Explicitly,
\begin{equation}\label{eq:regress}
    g(\theta) = \tilde{\Phi}^\epsilon(\theta) (\tilde{\Phi}^\epsilon)^\dagger \hat{g},
\end{equation}
i.e. a simple dot product where the $M$-dimensional vector $(\tilde{\Phi}^\epsilon)^\dagger \hat{g}$ can be precomputed. The absence of $\Lambda^\epsilon$ in Eq. \eqref{eq:regress} means that this expression is stable and does not suffer from instabilities. 

In order to make this work in practice the parameters $M$ and $\alpha$ need to be chosen with some care. If $M$ is too small, the expansion from Eq. \eqref{eq:mercer} is not converged; if it is too large then the calculation of the pseudo-inverse $(\tilde{\Phi}^\epsilon)^\dagger$ again becomes ill-conditioned. If $\alpha$ is too large, the basis functions $\varphi_k^\epsilon$ are too narrow and vice versa, yielding again either poor convergence of Eq. \eqref{eq:mercer} or conditioning problems for $(\tilde{\Phi}^\epsilon)^\dagger$ \cite{fasshauer}. We found that $\epsilon = 0.1$ with $\alpha$ between $1.8$ and $2.5$ and $M\sim N_n/4$ to yield satisfactory results. Furthermore, we have checked that adding some fraction of noise ($\sim 10^{-5}$, thus below the accuracy levels of the interpolant) to the input weights computed at the Halton nodes leads to completely negligible differences in the resulting interpolant. %The exact interpolation problem is replaced by an approximate \textit{regression} for determining the $M$ unknown coefficients $b_k^\epsilon$ \footnote{We still loosely refer to the resulting approximation as an `interpolant', though strictly speaking this is inadequate. Its values do not match those of the input weights at the Halton nodes to machine precision.}. Notably, the computation of Eq. \eqref{eq:regress} does not involve the matrix $\Lambda^\epsilon$ containing the small eigenvalues at any point, to which the authors of \cite{fasshauer} attribute its numerical stability. This procedure was then dubbed \textit{RBF-QR regression}, with QR referring to the QR decomposition of $\tilde{\Phi}^\epsilon$ and regression to the $M$-term truncation of Eq. \eqref{eq:mercer}, which leads to an overdetermined system of equations for the coefficients $b_k^\epsilon$ since $M<N_n$.

In the $D$-variable case (keeping $\epsilon$ the same for all dimensions) we can write the exponential of the sum as a product of exponentials, so that Eq. \eqref{eq:mercer} is generalized to 
\begin{eqnarray}\label{eq:mercermult}
    e^{-\epsilon^2(\Theta - \Theta')_m(\Theta - \Theta')_m}  \approx \sum_{\mathbf{k}=\mathbf{1}_D}^\mathbf{M} \lambda_\mathbf{k}^\epsilon \varphi_\mathbf{k}^\epsilon(\Theta) \varphi_\mathbf{k}^\epsilon(\Theta')
\end{eqnarray}
where we use multi-index notation, i.e. $\mathbf{1}_D = (1,1,\dots 1)$ while $\mathbf{k} = (k_1, k_2, \dots k_D)$ and 
\begin{eqnarray}
    \varphi_\mathbf{k}^\epsilon(\Theta) = \prod_{m=1}^D \varphi_{k_m}^\epsilon(\Theta_m) 
\end{eqnarray}
and $\lambda_\bb{k}^\epsilon$ is just the product of all the $\lambda_{k_m}^\epsilon$. The size of the eigenvalues is thus dictated by the sum $S_m = k_1 + \dots + k_m$ of the indices: 
\begin{eqnarray}
    \lambda_\mathbf{k}^\epsilon = \bigg(\frac{\alpha^2}{\alpha^2+\epsilon^2+\delta^2}\bigg)^{D/2}\bigg( \frac{\epsilon^2}{\alpha^2+\epsilon^2+\delta^2}\bigg)^{S_m-D}.
\end{eqnarray}
For this reason, it is natural to truncate Eq.  \eqref{eq:mercermult} at some value $S_\text{max}$ of $S_m$, thus keeping all terms with index sum at most $S_\text{max}$. From Eq. \eqref{eq:herm}, we infer that this corresponds to keeping all polynomials in the $\Theta_m$ variables of total degree $S_\text{max}-D$, of which there are $M = {S_\text{max} \choose D}$. 
\section{Curvature, Neutrinos and Dynamical Dark Energy}\label{sec:ext}

We now illustrate the power and generality of our approach by extending the treatment of the $\Lambda$CDM power spectrum to a much larger space of cosmological parameters. Specifically, we vary the neutrino mass $M_\nu$ while also considering the CPL dynamical dark energy parametrization $w(a) = w_0 + w_a(1-a)$ \cite{cpl1,cpl2} and nonzero curvature. The weights $w_i(\Theta)$ now depend on nine rather than just three parameters. While we no longer benefit from the separability of Eq. \eqref{eq:evol} and strictly speaking all parameters except $A_s$ affect the shape of the power spectrum, some of them still do so only mildly. It suffices to fix $A_s, h, z$ and $w_a$ for the purposes of constructing the templates, so that any change in the shape due to variations in the parameters can simply be absorbed into the weights. The associated ranges and grid sizes are indicated in Table \ref{tab:tab2}. 
We emphasize again that the templates do not need to be computed exactly, and we need at most $\sim 80\,000$ calls to \texttt{CAMB} for the SVD \footnote{We can mimic the effect of curvature on the shape on large scales by multiplying all spectra by an (empirical) `fudge factor' of the form $f(k) = 1+a_0 \Omega_K (k_0 / k)^2$ for a linear grid of $12$ values of $\Omega_K$. The dependence of the power spectrum on $n_s$ is trivial, meaning that templates for varying $n_s$ do not need to be recomputed either (as long as one works with a rectangular grid). An empirical correction may also be possible for the dark energy equation of state parameter $w_0$, further reducing the number of power spectrum evaluations necessary. We did not pursue this in more detail.}\footnote{For simplicity, we refrain from extending the range for $\Omega_K$ further since this would allow $\Omega_\Lambda<0$ in which case the expansion from Eq. \eqref{eq:curvgr} is no longer valid \cite{hamilton_growth}.}. For the extended range, we use the parameter $w_+=w_0+w_a$ rather than $w_a$ itself. Note that we need $w_+<0$ in order to enforce a period of early matter domination\footnote{We have attempted bringing $w_+$ closer to zero, but this resulted in significantly degraded performance, in line with findings of \cite{knabe2_v2}. In any case, this region of parameter space is quite pathological and appears to be disfavoured \cite{desibaov2}.}. With the possible exception of time-varying dark energy, the extended parameter ranges are broad and uninformative for Stage IV galaxy surveys \cite{chudaykin_mcmc,braganca,chudaykin_de}.
\begin{table}[t]
    \centering
    \begin{tabular}{|c|c|c|c|c|}
    \hline
         & \multicolumn{2}{c|}{Default} &
        \multicolumn{2}{c|}{Extended} \\
        \hline
        $\Theta $& Range & Grid size & Range & Grid size \\
        \hline\hline
        $\omega_c$ & [0.095,0.145] & 20 & [0.082,0.153] & 30  \\
        $\omega_b$ & [0.0202,0.0238] & 12 & [0.020,0.025] &12 \\
        $\Omega_K$ & [-0.12,0.12] &12 & [-0.2,0.2] &12 \\
        $h$ & [0.55,0.8] & $h^* = 0.7$ & [0.51,0.89] & $h^* = 0.7$\\
        $n_s$ & [0.9,1.02] &8 & [0.81,1.09]&15 \\
        $M_\nu$ [eV] & [0,0.6] &12 & [0,0.95] &15\\
        $w_0$ &[-1.25,-0.75] &12 & [-1.38,-0.62] &15\\
        $w_a/w_+$ &[-0.3,0.3] &$w_a^*=0$ &[-1.78,-0.42]&$w_+^*=-1$\\
        $z$ & [0.1,3] & $z^* = 0$ & [0.1,3] & $z^* = 0$ \\
        $10^9 A_s$ & - & $10^9 A_s^* = 2$ & - &$10^9 A_s^* = 2$ \\
        \hline
    \end{tabular}
    \caption{Ranges and template grids for the generalized cosmological parameter space. If a parameter is held fixed for the templates, its fiducial value is indicated instead. %The volume of the extended space is roughly $\sim 65$ times larger than that of the default space.
    }
    \label{tab:tab2}
\end{table}

Extracting the first $30$ components takes $\mathcal{O}(10)$ min on a single CPU (but requires significant memory if the number of templates is large), while determining the coefficients of the RBF approximation of the weights takes negligible time. We use $N_n = 20\,000$ Halton nodes across the $9$-dimensional parameter space. As for the $\Lambda$CDM power spectrum, we again found it beneficial to normalize the weights before applying the RBF (details are given below). We use $\sim 20\,000$ randomly selected cosmologies to test the precision of the predictions. 

\begin{figure}[t]
\includegraphics[width=0.48\textwidth]{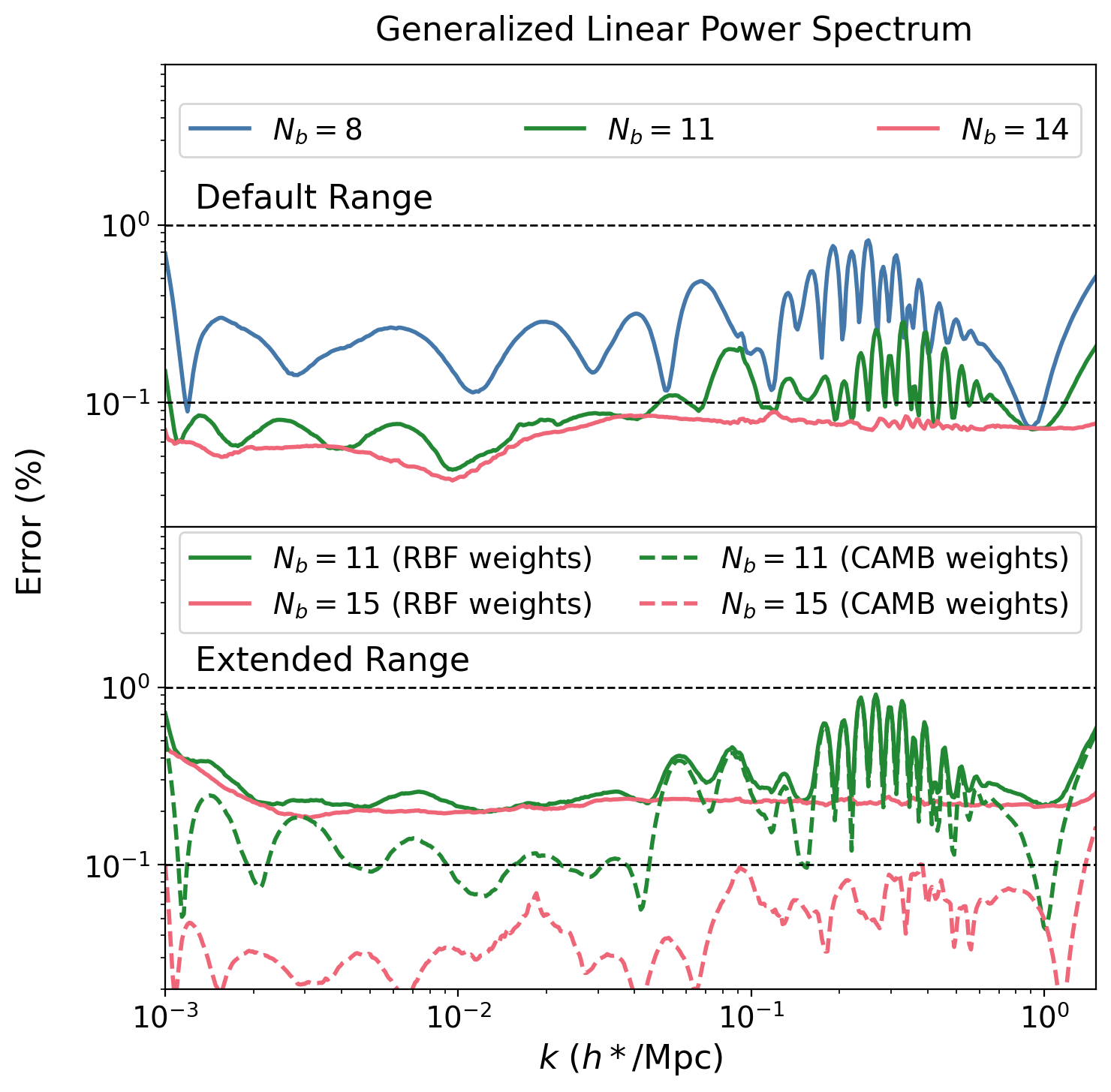}
\caption{\label{fig:plarge} The $99.7$th percentile of the absolute errors on the generalized linear power spectrum as a function of scale for several choices of number of basis functions used. The upper panel corresponds to the default range, while the lower panel corresponds to the extended range. Dashed black lines indicate $0.01\%,\, 0.1\%$ and $1\%$ errors, respectively. Dashed curves indicate the error if the weights are calculated \textit{exactly} (i.e. using \texttt{CAMB} and Eq. \eqref{eq:proj}), while solid lines indicate the error if they are calculated using RBFs.}
\end{figure}

The precision of the resulting linear power spectrum decomposition is shown in Figure \ref{fig:plarge}. 
We are again able to reach below $0.1\%$ precision for $99.7\%$ of the test samples for the default range and below $0.2\%$ for the extended range. This increase is due to inaccuracies in the RBF approximation, which is not surprising given that we have used the same number of Halton nodes for both the default and the extended parameter ranges. Increasing the number of Halton nodes will likely ameliorate this issue, but we did not pursue this any further. Importantly, the dashed curves in the bottom panel of Figure \ref{fig:plarge} illustrate that the accuracy of the decomposition from Eq. \eqref{eq:decomp} is still satisfactory - using e.g. $N_b=19$ would decrease the error further to $\sim 0.05\%$ across all scales. Calculating a single power spectrum takes $\sim 1$ ms, while calculating $250$ predictions takes $\sim 30$ ms. The speed-up from vectorization is thus less pronounced compared to the $\Lambda$CDM case.

In the remainder of this Appendix, we provide further details on how we interpolated the weights $w_i(\Theta)$ using RBFs. First, we found it beneficial to factor the calculation of the weights as
\begin{fleqn}
\begin{align}
    w_i(\Theta) = \frac{w_i(\Theta)}{\sqrt{P_L^\Theta(k_1)P_L^\Theta(k_2)}} \frac{\sqrt{P_L^\Theta(k_1)P_L^\Theta(k_2)}}{D^2_+(\Theta)} D^2_+(\Theta)
\end{align}
\end{fleqn}
where $D_+(\Theta)$ is the scale-independent growth factor in the corresponding cosmology \textit{without} neutrinos, $k_1 \approx 0.009h^*/$Mpc and $k_2 \approx 0.47h^*/$Mpc. The two ratios in the above expression are rather slowly varying functions of cosmology and can hence be approximated accurately using our RBF method. Thus, after applying the RBF approximation to these ratios (which still depend nontrivially on all nine parameters) it remains to compute $D_+(\Theta)$, which depends on six parameters in total (i.e. the total physical cold matter density $\omega_m = \omega_b+\omega_c$ as well as $\Omega_K,h,w_0,w_a$ and $z$). Solving the ordinary differential equation that defines $D_+$ would be slow compared to the calculation of the RBF interpolant and the sum over scale functions, and there is no hope of making analytic progress in the case of a time-varying equation of state $w_a \neq 0$. Therefore, we adopt a different procedure.

The growing mode in a universe with ordinary matter, cosmological constant and curvature can be expanded in $\Omega_K$ and is given (for positive $\Omega_\Lambda$ and $|\Omega_K/(1-\Omega_K)|<1$) by the series \cite{hamilton_growth}
\begin{eqnarray}\label{eq:curvgr}
    \frac{D_+(\Theta)}{a} &=& R(a)\sum_{n=0}^\infty \frac{5(-1)^n (3/2)_n}{n!(5+2n)}\bigg(\frac{\Omega_K q^{1/3}}{\Omega_m^{2/3}\Omega_\Lambda^{1/3}(1+q)}\bigg)^n \nonumber \\
    &\times& {}_2 F_1\left(1,\frac{1-2n}{3},\frac{11+2n}{6}, -q\right)
\end{eqnarray}
where $q =\Omega_\Lambda a^3/\Omega_m$ and $R(a) = H(a)/H_{K=0}(a)$ is the ratio of the Hubble factor to its value when the curvature term is set to zero, i.e.
\begin{eqnarray}\label{eq:ratio}
    R^2(a) = \frac{\Omega_m a^{-3} + \Omega_K a^{-2} + \Omega_\Lambda}{\Omega_m a^{-3} + \Omega_\Lambda}.
\end{eqnarray}
Furthermore, there also exists an exact solution for the growing mode in cosmologies with zero curvature, but non-trivial (constant) equation of state $w_0$ \cite{lee_growth}: 
\begin{equation}\label{eq:wsol}
    \frac{D_+(\Theta)}{a} =  {}_2 F_1\left(\frac{w_0-1}{2w_0},\frac{-1}{3w_0},1-\frac{5}{6w_0}, -q_w\right)
\end{equation}
with $q_{w} = \Omega_\Lambda a^{-3w_0}/\Omega_m $.
A solution in the case of non-vanishing curvature as well as non-trivial equation of state appears to be unknown, but we can instead make the following ansatz, denoted by an overbar to distinguish it from the true solution:
\begin{eqnarray}\label{eq:ansatzgr}
    \frac{\bar{D}(\Theta)}{a} &=& R(a)\sum_{n=0}^\infty \frac{5(-1)^n (3/2)_n}{n!(5+2n)}\bigg(\frac{\Omega_K q_w^{1/3}}{\Omega_m^{2/3}\Omega_\Lambda^{1/3}(1+q_w)}\bigg)^n \nonumber \\ 
    &\times & {}_2 F_1\left(\frac{w_0-1}{2w_0},\frac{2n-1}{3w_0},1-\frac{5+2n}{6w_0}, -q_w\right),
\end{eqnarray}
where now the time variable in the summation is
\begin{eqnarray}
    q_w = \frac{\Omega_\Lambda}{\Omega_m} a^{-3(w_0+w_a)}\exp\big( 3w_a(a-1)\big)
\end{eqnarray}
and the prefactor $R(a)$ is modified correspondingly by substituting \cite{cpl2}
\begin{eqnarray}
    \Omega_\Lambda \to \Omega_\Lambda a^{-3(1+w_0+w_a)}\exp\big( 3w_a(a-1)\big)
\end{eqnarray}
in Eq. \eqref{eq:ratio}.
The expression in Eq. \eqref{eq:ansatzgr} is an amalgamation of Eq. \eqref{eq:curvgr},\eqref{eq:wsol}. By construction, $\bar{D}$ reduces to the above exact solutions in the limiting cases where $w_a =0$ and either $\Omega_K=0$ or $w_0=-1$, respectively. We decomposed the ratio of the exact growth factor by this quantity, while keeping the first $15$ terms in the series expansion (this can be computed at negligible cost). Thus, the weights are factorized as 
\begin{eqnarray}\label{eq:finfac}
    w_i(\Theta) &=& \frac{w_i(\Theta)}{\sqrt{P_L^\Theta(k_1)P_L^\Theta(k_2)}} \frac{\sqrt{P_L^\Theta(k_1)P_L^\Theta(k_2)}}{D^2_+(\Theta)}\nonumber \\
    &\times &
    \frac{D^2_+(\Theta)}{\bar{D}^2(\Theta)}\bar{D}^2(\Theta).
\end{eqnarray}
Here, we use $15\,000$ Halton nodes in the six-dimensional parameter space corresponding to the third term in the above expression.

For the default parameter ranges from the left column of Table \ref{tab:tab2}, Eq. \eqref{eq:finfac} was sufficient for our purposes as it achieves an accuracy of below $0.1\%$ for $99.7\%$ of the test cosmologies (top panel of Figure \ref{fig:plarge}). However, for the extended parameter ranges from the right column of Table \ref{tab:tab2} it yielded an overall error just below $1\%$ on the test spectra, which we deemed unacceptable. We managed to mitigate this issue via the following two modifications: (i) instead of directly applying the RBF to the second factor in Eq. \eqref{eq:finfac}, we first applied a scaled logit transformation \footnote{see e.g. \url{https://docs.scipy.org/doc/scipy/reference/generated/scipy.special.logit.html}} and then transformed back, and (ii) instead applying the RBF to $D^2_+(\Theta)/\bar{D}^2(\Theta)$ we applied it to $D^2_+(\Theta)/\bar{D}^{2\alpha}(\Theta)$ with $\alpha = 0.35$. The result of this procedure is what is shown in the bottom panel of Figure \ref{fig:plarge} as the solid curves. These two modifications did not have any discernible impact on the results for the default range, so we did not apply them there. We emphasize again that there are many alternative procedures that could lead to similar or improved results for the calculation of the weights \footnote{See e.g. the fast ODE solver from \url{https://cosmologicalemulators.github.io/Effort.jl/dev/} based on \cite{odesolv1,odesolv2}. We thank Marco Bonici for pointing this out to us.}, but we leave these investigations as well as a thorough comparison to our RBF method for future work. 

\section{IR Resummation}\label{sec:irres}
\begin{figure}[t]
\includegraphics[width=0.47\textwidth]{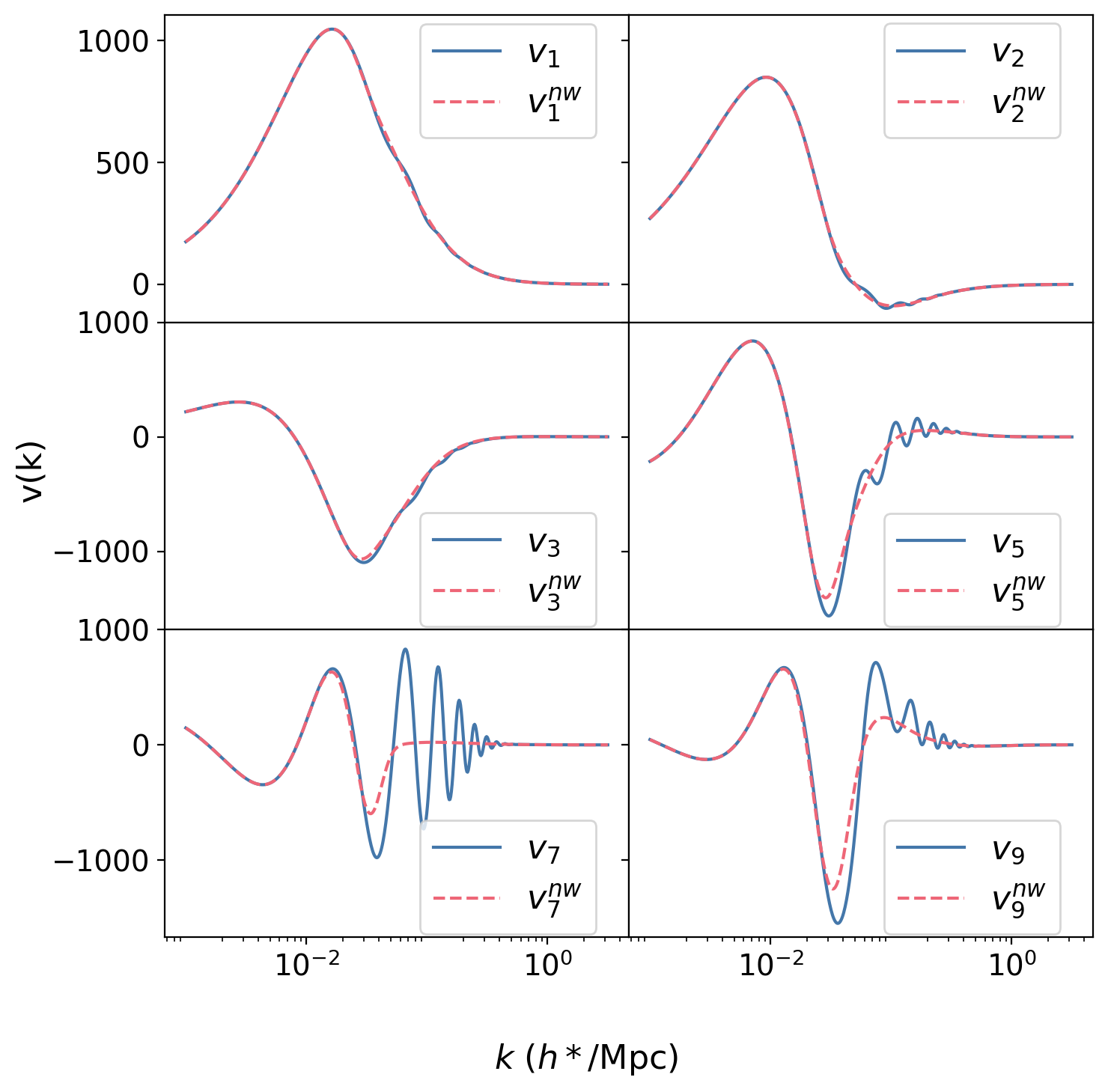}
\caption{\label{fig:irres} Some examples of scale functions $v_i(k)$(solid blue) along with their no-wiggle counterparts  $v_i^{\text{nw}}(k)$ (dashed pink)  obtained by applying Eq. \eqref{eq:dewig}. These correspond to the basis for the extended $\Lambda$CDM settings.}
\end{figure}

The observed galaxy power spectrum exhibits baryon acoustic oscillations with a smaller amplitude than the expectation from linear theory. This feature can be consistently taken into account via IR resummation. In Eulerian space, this is typically done by adopting a wiggle-no-wiggle split of the linear power spectrum. The no-wiggle part $P_{\text{nw}}$ is constructed by applying a suitable smoothing operation $\mathcal{F}$ to the linear power spectrum \cite{baldauf_bao,vlah_ir,ivanov_irres}: 
\begin{equation}\label{eq:wnw}
    P_{\text{nw}}^\Theta(k) = \mathcal{F}[P^\Theta_L](k).
\end{equation}
The IR-resummed power spectrum at the linear level in redshift space is then given by \cite{baldauf_bao,vlah_ir,ivanov_tspt2,ivanov_irres}
\begin{eqnarray}\label{eq:linir}
    P^\Theta_{L,\text{IR}}(k,\mu)&=&P_{\text{nw}}^\Theta(k)+e^{-k^2\Sigma_s^2(\Theta,\mu)}(P^\Theta_L(k)-P_{\text{nw}}^\Theta(k)) \nonumber \\
    &=&P_{\text{nw}}^\Theta(k)+e^{-k^2\Sigma_s^2(\Theta,\mu)}P_{\text{w}}^\Theta(k),
\end{eqnarray}
with the redshift-space damping factor,
\begin{eqnarray}\label{eq:irdamp}
    \Sigma_s^2(\Theta,\mu) = (1+f(\Theta)(f(\Theta)+2)\mu^2)\Sigma^2(\Theta),
\end{eqnarray}
and the isotropic real-space damping,
\begin{eqnarray}
    \Sigma^2(\Theta) = \int_0^{k_\text{o}}\frac{\md q}{6\pi^2}(1-j_0(qr_\text{o}) + 2j_2(qr_\text{o})P_\text{nw}(q).
\end{eqnarray}
We use $r_\text{o} = 105 \text{ Mpc}/h^*$ and $k_\text{o} = 0.2 h^*/\text{Mpc}$, independently of cosmology. At one-loop order the power spectrum becomes \cite{ivanov_tspt2,ivanov_irres}
\begin{eqnarray}\label{eq:1loopir}
    P^\Theta_{\text{1-loop,IR}}(k,\mu) &=& \text{const.} + \mathcal{S}^{l}[P^\Theta_{\text{nw}}]\nonumber \\ 
    &+& e^{-k^2\Sigma_s^2}(1+k^2\Sigma_s^2)\mathcal{S}^{l}[P^\Theta_{\text{w}}]\nonumber \\
    &+& \mathcal{S}^{q}[P^\Theta_{L,\text{IR}},P^\Theta_{L,\text{IR}}]
\end{eqnarray}
%We have explicitly indicated the possible cosmology dependence of the filter. 
where we suppressed some arguments to avoid clutter and $\mathcal{S}^l,\mathcal{S}^q$ are as in Eq. \eqref{eq:1loop1}. The last line of Eq. \eqref{eq:1loopir} depends on cosmology not only via the linear power spectrum, but also through the damping factor in the exponential from Eq. \eqref{eq:irdamp}. A commonly used form is to write the quadratic term as \cite{baldauf_bao,vlah_ir,ivanov_irres,egge_emu} 
\begin{eqnarray}\label{eq:irapprox}
    \mathcal{S}^{q}[P^\Theta_{L,\text{IR}},P^\Theta_{L,\text{IR}}] &\approx& \mathcal{S}^{q}[P^\Theta_{\text{nw}},P^\Theta_{\text{nw}}]\\
    &+&e^{-k^2\Sigma_s^2}\bigg(\mathcal{S}^{q}[P^\Theta_{L},P^\Theta_{L}]-\mathcal{S}^{q}[P^\Theta_{\text{nw}},P^\Theta_{\text{nw}}]\bigg). \nonumber
\end{eqnarray}
Various methods have been proposed in the literature to accomplish the smoothing in Eq. \eqref{eq:wnw} \cite{baldauf_bao,vlah_ir,lightrelics,chudaykin}. We would like the decomposition of Eq. \eqref{eq:decomp} to translate immediately into a decomposition of the wiggle and no-wiggle parts of a given power spectrum. The straightforward way to achieve this is by choosing a smoothing operation that is \textit{linear}, i.e. $\mathcal{F}[\lambda P] = \lambda \mathcal{F}[P]$ and $\mathcal{F}[P_1+P_2] = \mathcal{F}[P_1]+\mathcal{F}[P_2] $. Then, provided that we have a satisfactory approximation of the linear power spectrum, we can assert that 
\begin{eqnarray}\label{eq:nowigdec}
    P_{\text{nw}}^\Theta(k) &=& \mathcal{F}[w_i(\Theta)v_i](k) = w_i(\Theta) \mathcal{F}[v_i](k)\nonumber \\
    &=& w_i(\Theta) v_{i}^{\text{nw}}(k),
\end{eqnarray}
and correspondingly 
\begin{eqnarray}\label{eq:wigglydec}
    P_{\text{w}}^\Theta(k) &=& w_i(\Theta) (v_i(k) - \mathcal{F}[v_i](k)) \nonumber \\ 
    &=& w_i(\Theta) v_{i}^{\text{w}}(k).
\end{eqnarray}
In other words, if the filter $\mathcal{F}$ is linear, the scale functions $v_i(k)$ for the linear power spectrum can be smoothed individually to yield appropriate bases $v_{i}^{\text{w}}(k)$ and $v_{i}^{\text{nw}}(k)$ for the wiggle and no-wiggle parts, respectively. If the filter were explicitly cosmology dependent, then the separation of the cosmology dependence and scale dependence would no longer be realized already at the linear level. Hence, we need to choose a filter that is also not cosmology dependent. To this end, we employ the Gaussian 1D filter in logarithmic wavenumbers from \cite{vlah_ir}, after normalizing the linear power spectrum by the Eisenstein-Hu power spectrum $P^*_\text{EH}(k)$ \cite{ehspec} at a \textit{fixed} cosmology: 
\begin{eqnarray}\label{eq:dewig}
    P^\Theta_{\text{nw}}(k) &=& P^*_\text{EH}(k)\frac{1}{\lambda(k)\sqrt{2\pi}}\int \md \log{q}\frac{P^\Theta_L(10^{\log{q}})}{P^*_\text{EH}(10^{\log{q}})} \nonumber \\ 
    &\times & \exp{\bigg(-\frac{1}{2\lambda(k)^2}(\log{q}-\log{k})^2\bigg)}.
\end{eqnarray}
Here it is important to introduce a scale dependence in $\lambda$ so that it becomes small at both high and low $k$. This is done in order to make sure that the cutoffs introduced when computing the loop integrals (cf. Section \ref{sec:rsd}) do not spoil the agreement between $P_{\text{nw}}(k)$ and $P_L(k)$ at high and low values of $k$. The result of applying the filter to the scale functions is shown in Figure \ref{fig:irres}. For $i=1$, the scale function has essentially the same shape as the linear power spectrum and exhibits no zero crossings. However, the higher scale functions have more and more features. It is not at all clear from intuition what the smoothed version of all but the lowest few scale functions should look like; however, by virtue of the linearity of the filter we are able to ensure that no spurious features are introduced. 

\begin{figure}
\includegraphics[width=0.47\textwidth]{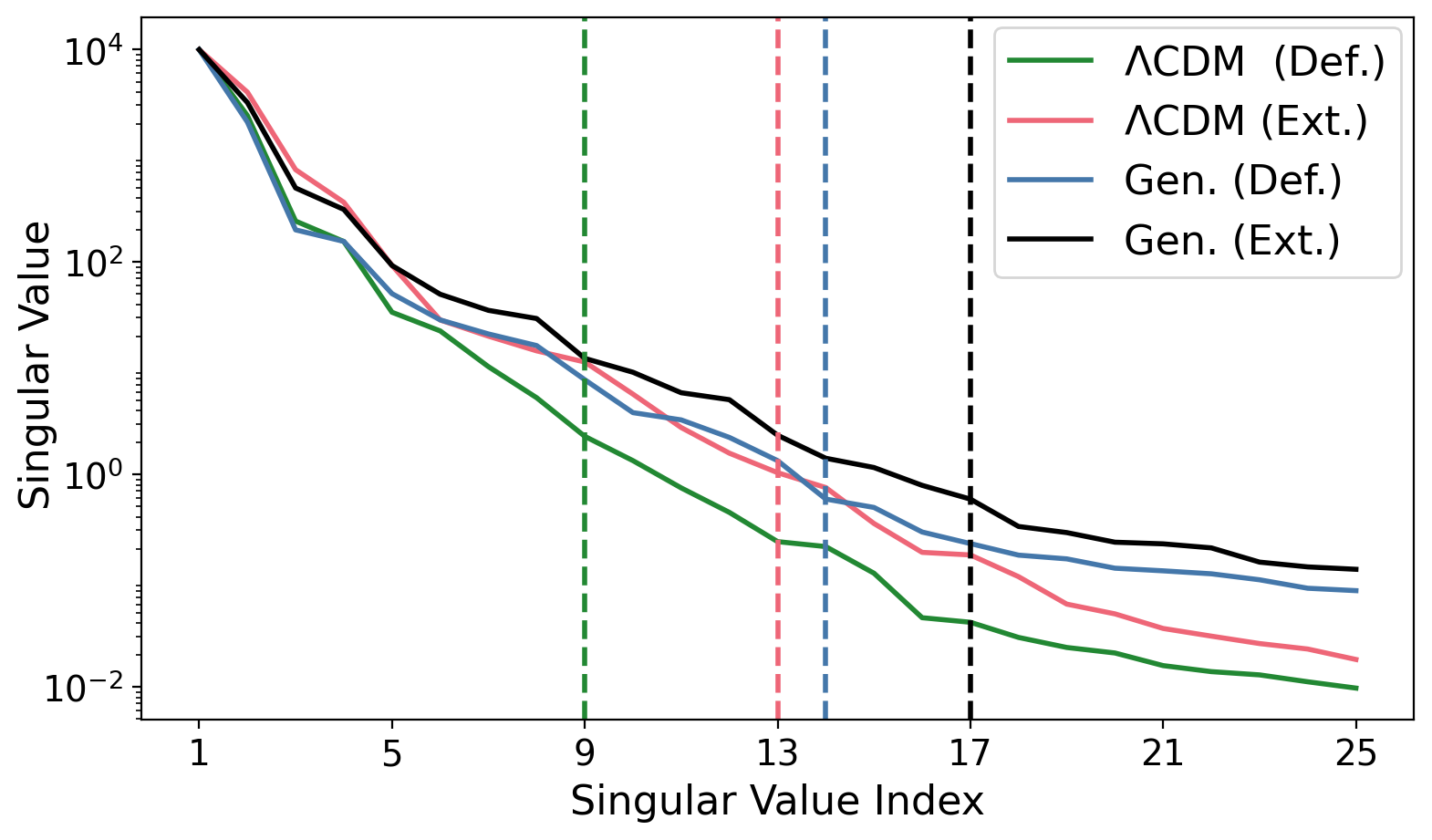}
\caption{\label{fig:svs} First 25 singular values of the SVD for all four parameter spaces considered in this work, arbitrarily normalized to have the same largest singular value. The horizontal lines indicate the number of basis functions needed to reach $\sim 0.1\%$ precision for $99.7\%$ of the test samples at $k\sim 0.1h^*$/Mpc.}
\end{figure}

In practice, the IR resummation procedure \textit{reduces} the error on the linear power spectrum around the BAO scale since wiggles are damped by an exponential factor and thus contribute less to the total (resummed) power spectrum. % with a known cosmology dependence.
This can be readily explained by looking at the behaviour of the wiggle and no-wiggle scale functions and the associated weights. The amplitude of the no-wiggle scale functions stays roughly the same, while the associated singular values (taken as a measure for the associated weights) decay quickly, as shown in Figure \ref{fig:svs}. Consequently, Eq. \eqref{eq:nowigdec} becomes accurate very quickly. Interestingly, however, the amplitude of the wiggly scale functions $v_i^\text{w}(k)$ clearly \textit{increases} with the index $i$, so that the convergence of the wiggle part given in Eq. \eqref{eq:wigglydec} is somewhat slower than the no-wiggle part. Thus, the higher scale functions contribute relatively little to the broadband power and mostly to $P_\text{w}(k)$. This is why the IR-resummed power spectrum is more easily approximated with a given number of scale functions than the original power spectrum. 

The fact that fewer basis functions are needed for IR-resummed expressions underscores the efficacy of \texttt{COBRA}: it also implies that the number of basis terms needed for the loop corrections is correspondingly small. This is an immediate consequence of \texttt{COBRA} taking full advantage of the analytic structure of the loop corrections (cf. Eq. \eqref{eq:1loopir},\eqref{eq:irapprox}), while direct emulation of the one-loop galaxy power spectrum (i.e. the individual terms multiplying bias coefficients) necessarily increases the number of cosmology dependent basis coefficients and thus constitutes a suboptimal expansion. As such, this approach could suffer from unnecessary precision or efficacy losses compared to the linear theory prediction.

An alternative approach (which we did not pursue) would be to perform the wiggle-no-wiggle split for each cosmology from the start using any filter, and then perform the SVD for the wiggle and no-wiggle power spectra separately. The advantage of that approach might be that one does not need a linear filter. Given the discussion above, it is not clear which approach will provide a more optimal basis, and we leave this for future work. 

Finally, owing to Eq. \eqref{eq:irapprox} the separation of scale dependence and cosmology dependence is maintained also at the one-loop level. We have 
\begin{eqnarray}
    P^\Theta_{\text{1-loop,IR}}(k,\mu) &=& \text{const.} + \mathcal{S}_i^{l,\text{nw}}w_i(\Theta) \nonumber \\ 
    &+& e^{-k^2\Sigma_s^2}(1+k^2\Sigma_s^2)\mathcal{S}_i^{l,\text{w}}w_i(\Theta)\nonumber \\
    &+& \mathcal{S}_{ij}^{q,\text{nw}}w_i(\Theta)w_j(\Theta) \nonumber \\
    &+&e^{-k^2\Sigma_s^2} \mathcal{S}_{ij}^{q,\text{w}}w_i(\Theta)w_j(\Theta),
\end{eqnarray}
where the linear pieces are $\mathcal{S}_i^{l,\text{nw}}= \mathcal{S}^{l}(v^\text{nw}_i)$ and $\mathcal{S}_i^{l,\text{w}} = \mathcal{S}^{l}(v^\text{w}_i)$ while the quadratic pieces are $\mathcal{S}_{ij}^{q,\text{nw}} = \mathcal{S}^{q}[v^\text{nw}_i,v^\text{nw}_j]$ and $\mathcal{S}_{ij}^{q,\text{w}} = \mathcal{S}^{q}[v_i,v_j] - \mathcal{S}^{q}[v^\text{nw}_i,v^\text{nw}_j]$. This is what is used in Section \ref{sec:rsd}. We approximate $\Sigma^2$ in a similar fashion as in Section \ref{sec:decomp} and the growth rate by first normalizing by the analytic solution for $\Omega_m + \Omega_\Lambda=1$. 

Regarding the extensions of the IR resummation to the higher N-point functions, an analogous expression to the one given in Eq. \eqref{eq:1loopir} has also been derived for bispectrum \cite{ivanov_irres}, which can be recast into a form analogous to Eq. \eqref{eq:irapprox} without loss of accuracy. As with the power spectrum, this latter form is particularly convenient for implementation in the \texttt{COBRA} framework. Recently, an alternative derivation of the bispectrum IR-resummation was presented in \cite{Chen:2024}, where a comprehensive resummation of long displacement fields was performed, including the derivation of IR suppression in the wiggle part. These results also demonstrate how IR resummation can be applied to higher N-point functions, ultimately leading to a similarly convenient wiggle-no-wiggle form. In conclusion, the \texttt{COBRA} framework is well-suited for the implementation of the IR resummation via the wiggle no-wiggle approximations even in higher N-point functions. 

\end{document}